\documentclass[12pt,a4j]{article}
\usepackage[dvips]{color,graphicx}
\usepackage{amsmath,enumerate}
\newcommand{\bm}[1]{\mbox{\boldmath{$#1 $}}}

\begin{document}
\begin{center}
{\LARGE\textbf{Varying-coefficient modeling via}}\\
{\LARGE\textbf{regularized basis functions}}\\
\end{center}
\begin{center}
{\large Hidetoshi Matsui$^1$,  Toshihiro Misumi$^2$ and Shuichi Kawano$^3$}
\end{center}

\begin{center}
\begin{minipage}{14cm}
{
\begin{center}
{\it {\footnotesize 
$^1$ Nikon Systems Inc., \\
1-6-3, Nishiohi, Shinagawa-ku, Tokyo 140-0015, Japan,\\

\vspace{1.2mm}

$^2$ Astellas Pharma Inc., \\
3-17-1, Hasune, Itabashi-ku, Tokyo 174-8612, Japan, \\

\vspace{1.2mm}

$^3$ Department of Mathematical Sciences, Osaka Prefecture University,\\
1-1 Gakuen-cho, Sakai, Osaka 599-8531, Japan. \\
}}

\vspace{2mm}

{\small h.matsui@nikon-sys.co.jp (H. Matsui) \\
toshihiro.misumi@jp.astellas.com (T. Misumi) \\
skawano@ms.osakafu-u.ac.jp (S. Kawano)}
\end{center}
\vspace{1mm} 

{\small {\bf Abstract:} \
We address the problem of constructing varying-coefficient models based on basis expansions along with the technique of regularization. 
A crucial point in our modeling procedure is the selection of smoothing parameters in the regularization method. 
In order to choose the parameters objectively, we derive model selection criteria from the viewpoints of information-theoretic and Bayesian approach. 
We demonstrate the effectiveness of proposed modeling strategy through Monte Carlo simulations and analyzing a real data set. 
}

\vspace{3mm}

}
\end{minipage}
\end{center}

\section{Introduction}
Longitudinal data are encountered in various fields, e.g., medical research, economic science and so on.
In the setting of longitudinal study, the outcome data are measured repeatedly over time for each individual. 
Nowadays many modeling strategies have been studied for analyzing longitudinal data, both in parametric and nonparametric way (see, for example, Diggle {\it et al.}, 1994; Vonesh and Chinchilli, 1997).  

Under linear parametric modeling for longitudinal data, a linear mixed-effect model is widely used in several literatures (see, for example, Verbeke and Molenberghs, 2000).
The advantage of linear mixed-effect model is easy to handle the unbalanced data, which are highly occurred in the longitudinal study.
Meanwhile, nonparametric regression (Ruppert {\it et al.}, 2003) and functional data analysis (Ramsay and Silverman, 2005) has come to the front recently for the nonparametric approach toward longitudinal data analysis.  

They can capture the complex structure in the longitudinal data effectively. 
While these parametric or nonparametric approaches are very useful, there are some problems about the adequacy of the model assumptions and the potential impact of model mis-specifications on the analysis, which is especially arisen in parametric models (Hoover {\it et al.}, 1998).
In addition, it is not unusual that covariates may depend on time progresses.
Nevertheless, these approaches does not necessarily consider studying an association between covariates and a response with time.

One of the most useful model to overcome this problem is the varying-coefficient model (VCM).
Hastie and Tibshirani (1993) proposed the smoothing spline method for estimating VCMs.
The essential idea behind the VCM is that coefficients of regression model are represented as time-dependent function.
It enables us to study the association between the time varying covariates and outcome.
Hoover {\it et al.} (1998) presented two types of nonparametric estimation procedure for VCMs, smoothing spline and locally weighted polynomials. 
They used a cross validation for selecting smoothing parameters in smoothing spline method.
However, the cross validation requires large computational time and yields the high variability since the selector is repeatedly applied.

In this paper, we introduce a nonlinear varying-coefficient modeling strategy using a linear combination of basis functions and regularized likelihood estimation method for continuous longitudinal data.
We also note that adjusted parameters included in our proposed model are regularization parameters.
In order to choose these parameters, we derive model selection criteria from information-theoretic and Bayesian viewpoints.
The proposed nonlinear varying coefficient modeling procedure is investigated through the analysis of real data and Monte Carlo simulations.

The article is organized as follows.
In Section 2 and 3, we present the varying coefficient model based on basis expansion and its estimation by the maximum penalized likelihood method.
Section 4 provides model selection criteria derived from the information-theoretic and Bayesian approach.
In Section 5 we describe Monte Carlo simulations in order to examine the effectiveness of our modeling procedure, and then we also apply the proposed method to Multicenter AIDS Cohort Study data in Section 6.
Summary and discussion are given in Section 7.

\section{Varying-coefficient models}
Suppose we have $p$ sets of predictors $X_k$ $(k = 1,\ldots,p)$ and a response varying with time $Y(t)$, and denote $i$-th observations at time points $j = 1,\ldots, n_i$ as $x_{ijk}$, and $y_{ij}$, respectively.  Then the varying-coefficient model has the form (Hoover {\it et al.}, 1998)
\begin{eqnarray}
y_{ij} = \beta_0(t_{ij}) + x_{ij1}\beta_1(t_{ij}) + \cdots + x_{ijp}\beta_p(t_{ij}) + \varepsilon_{ij}, \label{varcoef1}
\end{eqnarray}
where $\beta_0(\cdot), \beta_1(\cdot), \ldots, \beta_p(\cdot)$ are parameter functions and $\varepsilon_{ij}$ are random noises whose vector $\bm\varepsilon_i = (\varepsilon_{i1},\ldots, \varepsilon_{in_i})'$ are normally distributed with mean vector $\bm 0$ and a variance covariance matrix $\Sigma_i$.  

We assume that coefficient functions $\beta_0(\cdot), \beta_1(\cdot), \ldots, \beta_p(\cdot)$ are expressed by basis expansions as follows;
\begin{eqnarray*}
\beta_k(t_{ij})=\sum_{m=1}^{M_k}\gamma_{km}\phi_m^{(k)}(t_{ij}) = \bm\gamma_k'\bm\phi^{(k)}(t_{ij}),
\end{eqnarray*}
where $\bm\gamma_k = (\gamma_{k1}, \ldots, \gamma_{kM_k})'$ are parameters to be estimated and $\bm\phi^{(k)}(t_{ij}) = (\phi_k^1(t_{ij}), \ldots, \\ \phi_k^{M_k}(t_{ij}))^{\prime}$ are basis functions.  There are various kinds of basis functions such as radial basis functions (Bishop, 1995; Ando {\it et al.}, 2008) or wavelets (Donoho and Johnstone, 1994; Fujii and Konishi, 2006). In this paper we apply $B$-spline bases.  Details of $B$-splines are referred to de Boor (2001) and Imoto and Konishi (2003).

Using the above assumption and denoting $\bm y_i = (y_{i1},\ldots, y_{in_i})'$, $D_{ik} = {\rm diag} (x_{i1k},\ldots, x_{in_i k})$ and $\Phi_{ik} = (\bm\phi^{(k)}(t_{i1}),\ldots, \bm\phi^{(k)}(t_{in_i}))'$, the varying coefficient model (\ref{varcoef1}) is rewritten as 
\begin{eqnarray}
\bm y_i = \sum_{k=0}^p D_{ik}\Phi_{ik}\bm\gamma_k + \bm\varepsilon_i,  \qquad \bm\varepsilon_i \sim N_{n_i}(\bm 0, \Sigma_i).
\label{varcoef2}
\end{eqnarray}
Then the varying coefficient model has the statistical model
\begin{align}
f(Y | \bm\theta) &= \prod_{i=1}^{n}\frac{1}{(2\pi)^{n_i/2}|\Sigma_i|^{1/2}} \nonumber \\
&\hspace{10mm} \times  \exp\left\{ -\frac{1}{2}\left(\bm y_i - \sum_{k=0}^p D_{ik}\Phi_{ik}\bm\gamma_k\right)'\Sigma_i^{-1}\left(\bm y_i - \sum_{k=0}^p D_{ik}\Phi_{ik}\bm\gamma_k\right)
\right\}, 
\end{align}
where $\bm\theta$ is a vector of unknown parameters.

\section{Estimation}
Unknown parameters, such as coefficient vectors $\bm\gamma_k$ and variance covariance matrices $\Sigma_i$ are estimated by the maximum penalized likelihood method, that is, maximizing the penalized likelihood defined by
\begin{eqnarray}
l_\lambda(\bm\theta) = l(\bm\theta)-\frac{n}{2}\sum_{k=1}^p \lambda_k\bm\gamma'_k\Omega_k\bm\gamma_k, \label{plike}
\end{eqnarray}
where $l(\bm\theta)$ is a log-likelihood function given by
\begin{align}
l(\bm\theta) &= \log f(Y|\bm\theta)\nonumber \\
&=-\frac{\sum_i n_i}{2}\log(2\pi) - \frac{1}{2}\sum_{i=1}^n \log|\Sigma_i| \nonumber\\
&\hspace{10mm} - \frac{1}{2}\sum_{i=1}^n\left(\bm y_i - \sum_{k=0}^p D_{ik}\Phi_{ik}\bm\gamma_k\right)'\Sigma_i^{-1}\left(\bm y_i - \sum_{k=0}^p D_{ik}\Phi_{ik}\bm\gamma_k\right)
\end{align}
and $\Omega_k$ is a positive semi-definite matrix.  Moreover, $\lambda_k$ are regularization parameters which control the effectiveness of the regularization.

Since it is difficult to derive estimates of parameters analytically, we apply the backfitting algorithm (Hastie and Tibshirani, 1990; 1993) for maximizing (\ref{plike}).  The first derivative of $l_\lambda(\bm\theta)$ with respect to $\bm\gamma_k$ is given by
\begin{align}
\frac{\partial l_\lambda(\bm\theta)}{\partial\bm\gamma_k}
&= \sum_{i=1}^n \left\{\Phi_{ik}'D_{ik}'\Sigma_i^{-1}\left(\bm y_i - \sum_{k=0}^p D_{ik}\Phi_{ik}\bm\gamma_k\right)\right\} - n\lambda_k\bm\gamma_k\Omega_k \nonumber \\
&= \sum_{i=1}^n \left\{\Phi_{ik}'D_{ik}'\Sigma_i^{-1}\left(\bm y_i  - \sum_{l\neq k}^p D_{il}\Phi_{il}\bm\gamma_l\right)\right\} \nonumber \\
&\hspace{20mm}  - \sum_{i=1}^n \Phi_{ik}'D_{ik}'\Sigma_i^{-1}D_{ik}\Phi_{ik}\bm\gamma_k- n\lambda_k\Omega_k\bm\gamma_k.
\end{align}
When coefficient parameters other than $k$-th are given, the backfitting algorithm iteratively estimates the $k$-th coefficient as follows: 
\begin{eqnarray}
\hat{\bm\gamma}_k = \left(\sum_{i=1}^n \Phi_{ik}'D_{ik}'\Sigma_i^{-1}D_{ik}\Phi_{ik} + n\lambda_k\Omega_k\right)^{-1}\left\{\sum_{i=1}^n \Phi_{ik}'D_{ik}'\Sigma_i^{-1}\left(\bm y_i - \sum_{l\neq k}^p D_{il}\Phi_{il}\hat{\bm\gamma}_l\right)\right\}
\end{eqnarray}
for $k = 1,\ldots, p$.  When variance covariance matrices $\Sigma_i$ are given in the form of $\Sigma_i = \sigma^2 S_i$ with an unknown parameter $\sigma^2$ and known matrices $S_i$, the parameter $\sigma^2$ is estimated by
\begin{eqnarray}
\hat\sigma^2 = \frac{1}{n}\sum_{i=1}^n\left(\bm y_i - \sum_{l\neq k}^p D_{il}\Phi_{il}\hat{\bm\gamma}_l\right)'S_i^{-1}\left(\bm y_i - \sum_{l\neq k}^p D_{il}\Phi_{il}\hat{\bm\gamma}_l\right).
\end{eqnarray}
Then we have the statistical model
\begin{align}
f(Y | \hat{\bm\theta}) &= \prod_{i=1}^{n}\frac{1}{(2\pi\hat{\sigma}^2)^{n_i/2}|S_i|^{1/2}} \nonumber \\
&\hspace{10mm} \times \exp\left\{-\frac{1}{2}\left(\bm y_i - \sum_{k=0}^p D_{ik}\Phi_{ik}\hat{\bm\gamma}_k\right)'\hat\sigma^{-2} S_i^{-1}\left(\bm y_i - \sum_{k=0}^p D_{ik}\Phi_{ik}\hat{\bm\gamma}_k\right)
\right\}.\label{statmodel}
\end{align}

\section{Model selection criteria}
The varying-coefficient model estimated by the maximum penalized likelihood method depends on regularization parameters. Smaller value of them yields overfitted estimates for the data, while the larger value provides models which does not capture the structure.  Therefore it is important to select appropriate values of them.  We consider using model selection criteria derived from the information-theoretic and Bayesian approach.  

Konishi and Kitagawa (1996) derived an information criterion GIC for evaluating models given by the M-estimate including maximum penalized likelihood method.  Using this result, the GIC for evaluating the varying-coefficient model estimated by the maximum penalized likelihood  described above is given by
\begin{eqnarray}
{\rm GIC} = -2l(\hat{\bm\theta}) + 2{\rm tr}\left\{R^{-1}(\hat{\bm\theta})Q(\hat{\bm\theta})\right\},
\end{eqnarray}
where $R(\bm\theta)$ and $Q(\bm\theta)$ are, respectively, given by
\begin{align}
R(\hat{\bm\theta})
= -\frac{1}{n}\sum_{i=1}^{n}\left.\frac{\partial^{2}\{l_\lambda^{(i)}(\bm{\theta})\}}{\partial\bm{\theta}\partial\bm{\theta}^{T}}\right|_{\theta = \hat\theta},~~
Q(\hat{\bm\theta})
= \frac{1}{n}\sum_{i=1}^{n}\left.\frac{\partial\{l_\lambda^{(i)}(\bm{\theta})\}}{\partial\bm{\theta}}\frac{\partial\{l^{(i)}(\bm{\theta})\}}{\partial\bm{\theta}^{T}}\right|_{\theta = \hat\theta},\label{RQ}
\end{align}
where $l^{(i)}_\lambda(\bm\theta) = l^{(i)}(\bm\theta) - (1/2)\sum_{j=1}^p \lambda_j\bm\gamma'_j\Omega_j\bm\gamma_j$ with the log-likelihood function of the $i$-th subject $l^{(i)}(\bm\theta)$.  The detailed elements of $R(\bm\theta)$ and $Q(\bm\theta)$ are given in the Appendix.

Konishi {\it et al.} (2004) extended the Schwarz's BIC (Schwarz, 1987) and derived a model selection criterion GBIC for evaluating models estimated by the penalized maximum likelihood method.  The GBIC for evaluating the varying coefficient model (\ref{statmodel}) is given by
\begin{align}
{\rm GBIC} =& -2f(Y|\hat{\bm\theta}) + n\sum_{k=1}^{p} \lambda_k\hat{\bm\gamma}'_k\Omega_k\hat{\bm\gamma}_k - \left(\sum_{k=1}^p r_k + 1\right)\log(2\pi)  \nonumber \\ 
&+\left(\sum_{k=1}^p r_k + 1\right)\log n 
- \sum_{k=1}^p \log|\Omega_k| + \log|R(\hat{\bm\theta})|, \label{GBIC}
\end{align}
where $r_k = M_k - {\rm rank}(\Omega_k)$.  The derivation of (\ref{GBIC}) is given in the Appendix.  We select a set of regularization parameters $\{\lambda_j\}$ which minimizes the value of either of these criteria and then consider the corresponding model as the optimal model.

\section{Monte Carlo simulations}
We conducted Monte Carlo simulations to examine the effectiveness of our modeling procedure. 
In the simulation study, we generated a data set $\{ (t_{ij}, y_{ij}, {\bm x}_{ij}) ; \ i=1,\ldots,n, \ j=1,\ldots,n_i \}$, where ${\bm x}_{ij} = (x_{ij1}, x_{ij2})^{\prime}$, given in the following. 
First, time points $t_{ij}$ were generated by $t_{ij} \sim U (0,1)$. 
Second, a response $y_{ij}$ and two predictors ${\bm x}_{ij}$ were derived as follows. 
\begin{eqnarray*}
&& y_{ij} = f(t_{ij}) + \varepsilon_{ij}, \quad f(t_{ij}) = x_{ij1} \beta_1 (t_{ij}) + x_{ij2} \beta_2 (t_{ij}), \\
&& x_{ij1} = a_i \cos (\pi t_{ij}) + b_i, \quad a_i \sim N(0,4), \quad b_i \sim U(2,3), \\
&& x_{ij2} = \{ 0, 1 \}, \\
&& \beta_1(t_{ij}) = \sin (\pi t_{ij}), \quad \beta_2 (t_{ij}) = t_{ij}, \\
&& \varepsilon_{ij} \sim N(0,\sigma^2), \quad \sigma = 0.05 \left\{ \max_{t \in [0,1]} f(t) - \min_{t \in [0,1]} f(t) \right\}.
\end{eqnarray*}
We considered four patterns of sample sizes; i.e., $n=10, 25, 50, 100$, and also $n_i$ was generated as an integer value between 8 and 15 for different suffix number $i$.

Based on the data set, we constructed an our varying-coefficient modeling procedure, where we use one-order (linear) $B$-splines as basis functions, a positive semi-definite matrix $\Omega_k \ (k=1,2)$ are assumed to be an identity matrix and the number of basis functions $M_k = \max_{i=1,\ldots,n} \{ n_i \}$ for $k=1,2$. 
Regularization parameters in penalized likelihood function were selected by the GIC or the GBIC. 
In order to investigate the efficiency of proposed modeling strategy, we compare the five-fold cross validation (CV), which is one of the most widely used in smoothing parameter selection, with the GIC and the GBIC. 
We repeated the procedure for 1000 times, and then obtained 1000 averages of mean squared errors ${\rm AMSE} = \sum_i \sum_j \{ f (t_{ij}) - \hat{y}_{ij} \}^2/(n \sum_i n_i)$, where $\hat{y}_{ij}$ is a predictive value.

\begin{table}[t]
\begin{center}
\caption{Comparisons of averaged mean squared errors using 1000 repetitions. }
\vspace{3mm}
\begin{tabular}{cccccccc}
\hline
$n=10$ & GIC & GBIC & CV & $n=25$ & GIC & GBIC & CV \\
\hline
AMSE & 36.45 & 36.48 & 36.59 & 
AMSE & 37.52 & 37.53 & 37.55\\
$\lambda_1 \ (\times 10^{-1})$ & 37.10 & 7.765 & 9.063 & 
$\lambda_1 \ (\times 10^{-1})$&34.74 & 9.574 & 10.49\\
$\lambda_2 \ (\times 10^{-2})$ & 1.680 & 1.683 & 1.506 & 
$\lambda_2 \ (\times 10^{-2})$&1.077 & 1.078 & 1.084\\
\hline
$n=50$ &  &  & & $n=100$ & & &\\
\hline
AMSE & 37.75 & 37.75 & 37.76 & 
AMSE & 38.15 & 38.15 & 38.16\\
$\lambda_1 \ (\times 10^{-1})$ & 34.85 & 10.94 & 13.83 & 
$\lambda_1 \ (\times 10^{-1})$ & 37.44 & 12.29 & 20.12\\
$\lambda_2 \ (\times 10^{-2})$ & 1.071 & 1.073 & 1.076 &
$\lambda_2 \ (\times 10^{-2})$ & 1.066 & 1.067 & 1.068\\
\hline
\end{tabular}
\label{table}
\end{center}
\end{table}

Table \ref{table} shows results of simulation studies and means of regularization parameters $\lambda_1$ and $\lambda_2$ for 1000 trials. 
It may be seen from the table that the models evaluated by the GIC or the GBIC are competitive or superior to those by the CV with respect to minimizing AMSE. 
Especially, when the sample size is small, our proposed methods seem to outperform methods by the CV.

\begin{figure}[t]
\begin{tabular}{cc}
\begin{minipage}{0.5\hsize}
\begin{center}
\includegraphics[height=6cm,width=7.5cm]{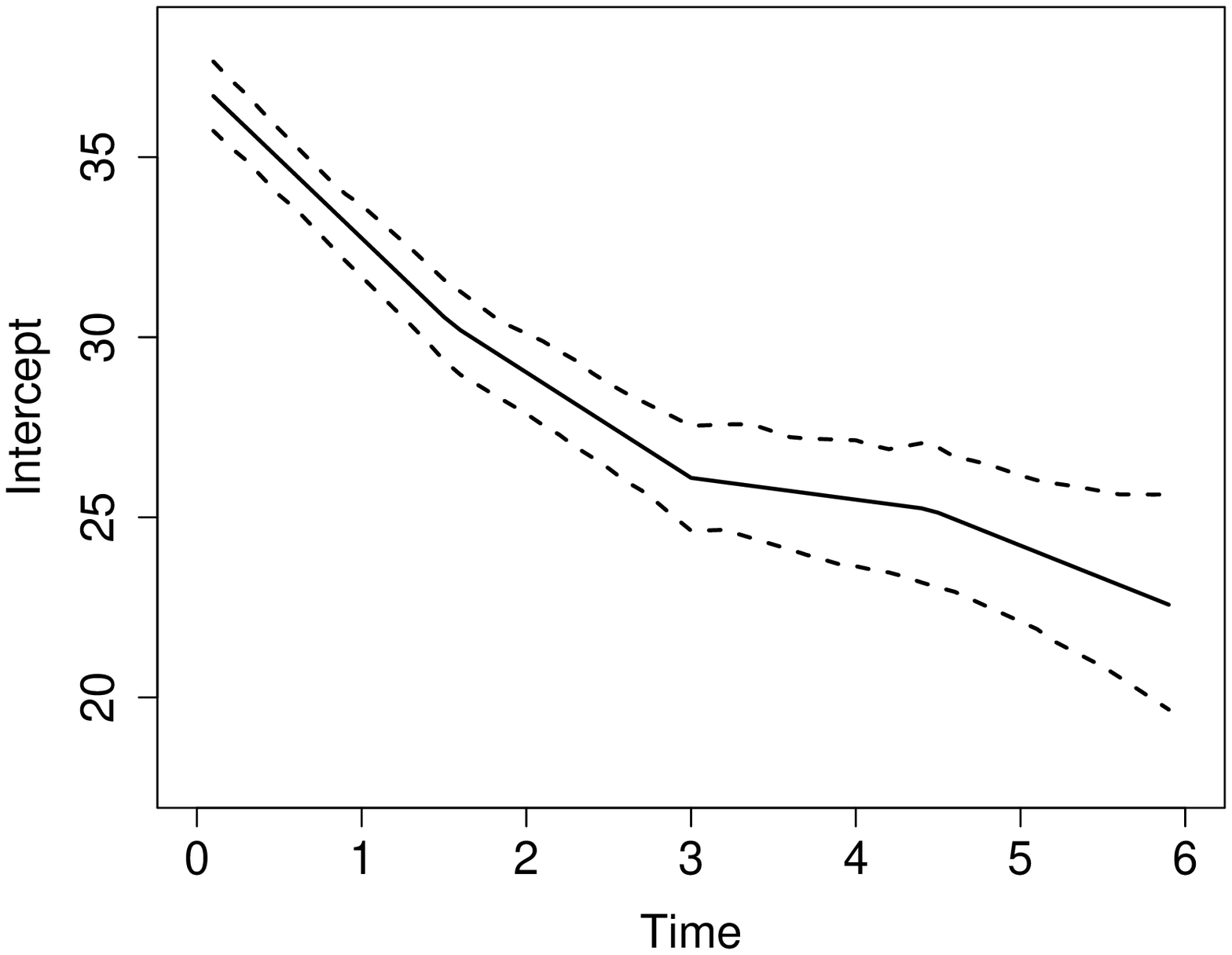}
\end{center}
\end{minipage}
\begin{minipage}{0.5\hsize}
\begin{center}
\includegraphics*[height=6cm,width=7.5cm]{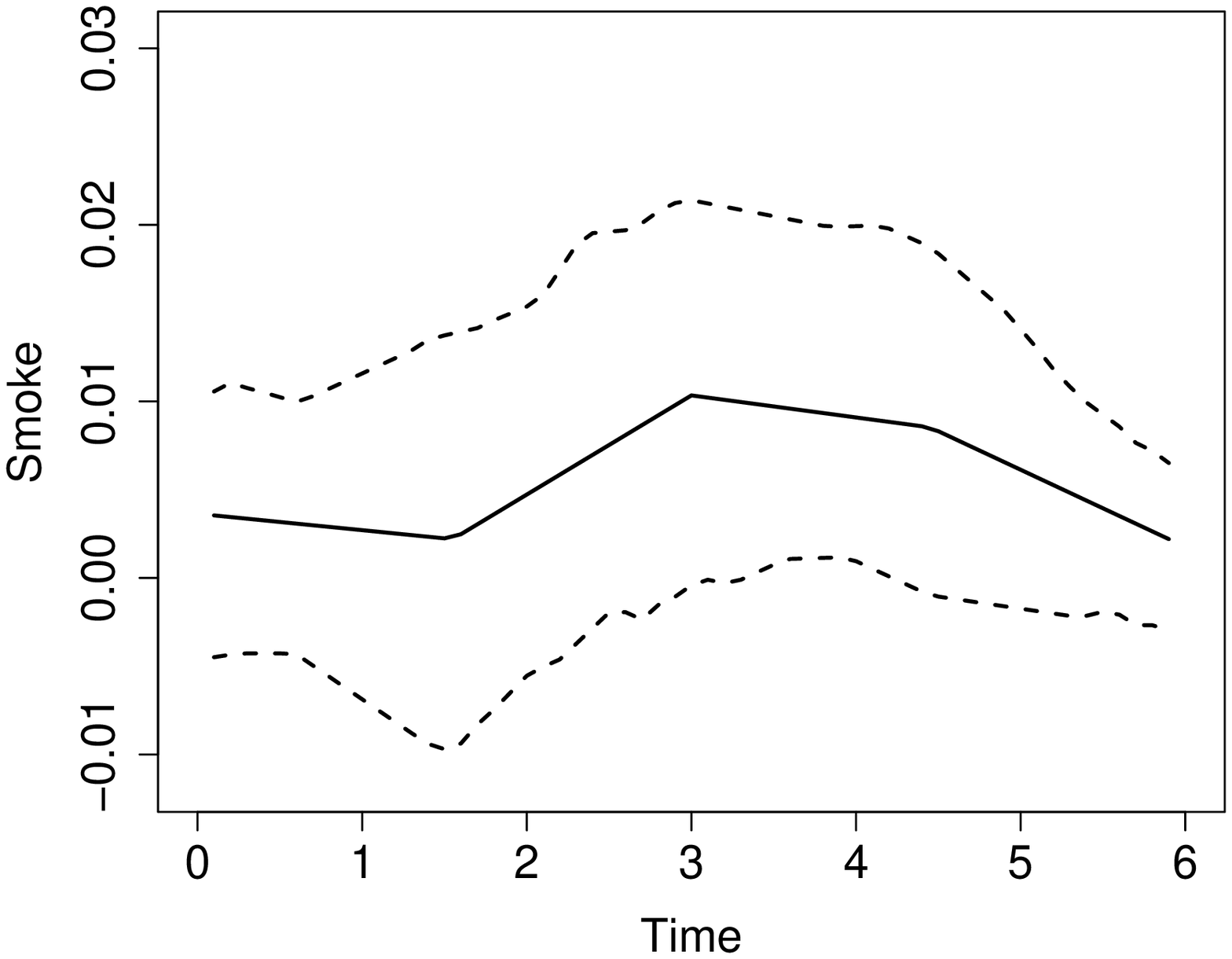}
\end{center}
\end{minipage}\\
\end{tabular}
\begin{tabular}{cc}
\begin{minipage}{0.5\hsize}
\begin{center}
\includegraphics*[height=6cm,width=7.5cm]{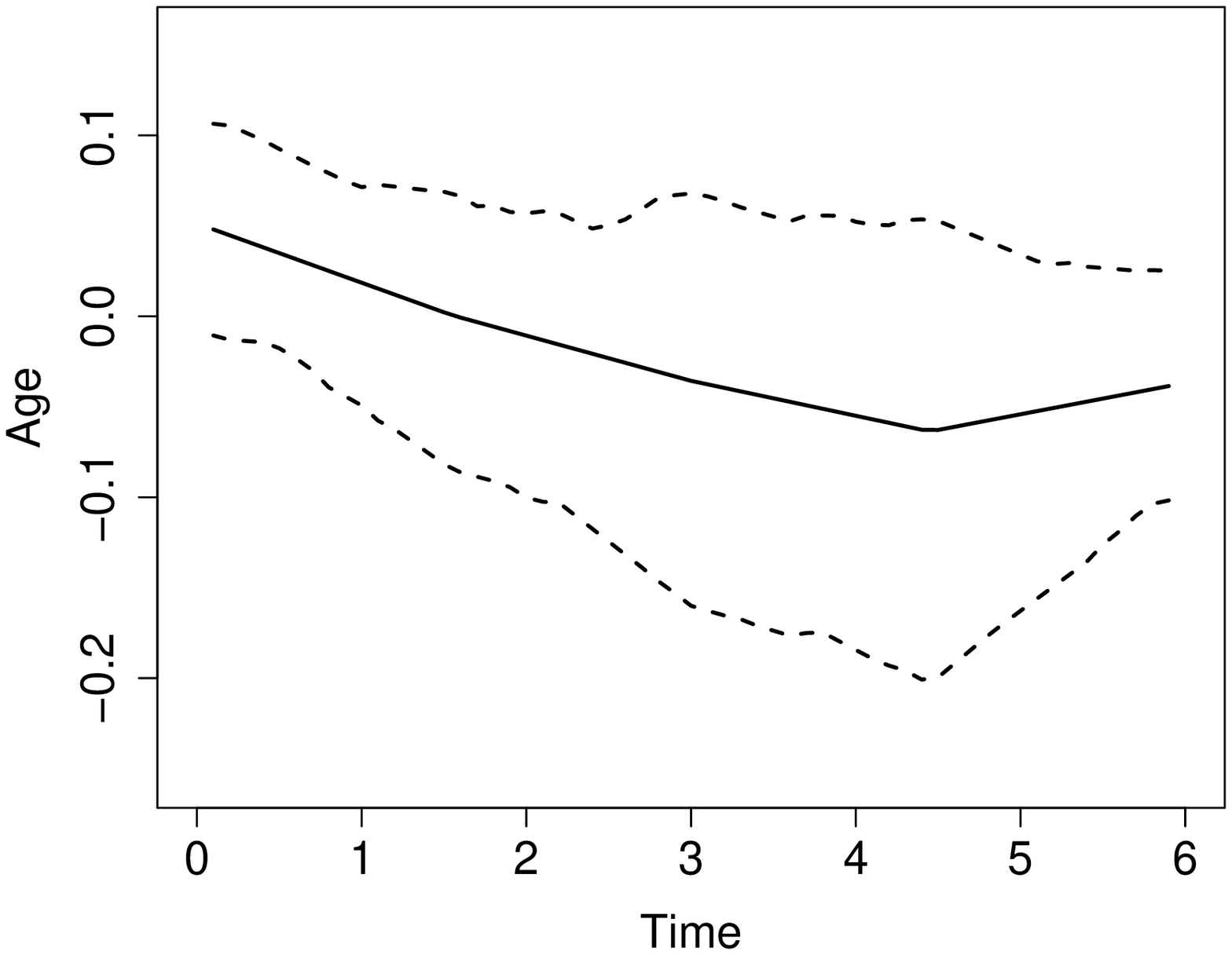}
\end{center}
\end{minipage}
\begin{minipage}{0.5\hsize}
\begin{center}
\includegraphics*[height=6cm,width=7.5cm]{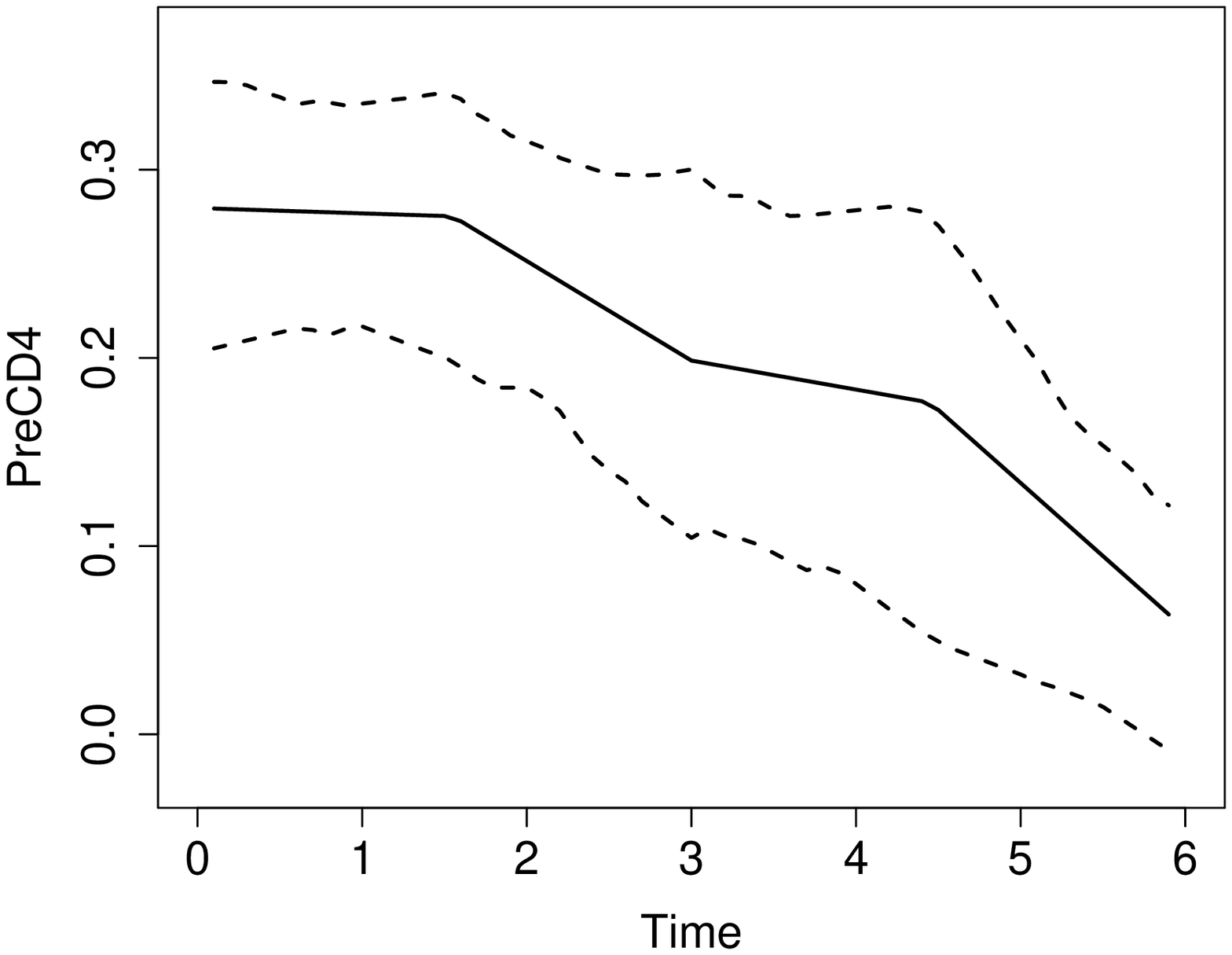}
\end{center}
\end{minipage}
\end{tabular}
\caption{
Estimated coefficient functions of the varying coefficient model. Top left: intercept, top right: smoking status, bottom left: age at HIV infection, bottom right: pre-infection CD4 percent.
}
\label{fig:coeff}
\end{figure}
\section{Real data example}
We applied the proposed modeling strategy to the analysis of the Multicenter AIDS Cohort Study data in order to capture the fluctuation of the percentage of the CD4 cells in the blood of the human who are infected with the Human Immunodeficiency virus (HIV).  
The data set contains cigarette smoking status, age at HIV infection, pre-HIV infection CD4 cell percent and the CD4 cell percentage of a subject, observed at distinct time points after HIV infection.  
Fan and Zhang (2000) analyzed them using functional version of the ANOVA models, and Huang {\it et al.} (2004) applied the time varying-coefficient models and then evaluated the model via the cross-validation.  

We represent the relationship of variables described above by the time varying-coefficient model written by
\begin{eqnarray*}
Y_i(t) = \beta_0(t) + X_{i1}\beta_1(t) + X_{i2}\beta_2(t) + X_{i3}\beta_3(t) + \varepsilon_i(t),
\end{eqnarray*}
where $X_{i1}$, $X_{i2}$, $X_{i3}$ represents centered cigarette smoking status, age at HIV infection and pre-infection CD4 percent of the $i$-th subject, respectively, $Y_i(t)$ denotes the CD4 percent of the $i$-th subject observed at differing time points, $\beta_j(t)$ $(j=0,\ldots, 3)$ are time varying coefficients and $\varepsilon_i(t)$ is the error function.  The model was fitted by the maximum penalized likelihood method with linear $B$-splines, and then it was evaluated by model selection GBIC derived in the previous section.  We generated 100 sets of bootstrap samples from the data, then obtained each estimates of coefficient functions.   

Figure \ref{fig:coeff} shows the result of the application of the varying coefficient modeling.  Solid lines are mean coefficient functions of 100 bootstrap samples and dashed lines are pointwise 95 \% confidence intervals.  The results suggest as follows. (1) The CD4 data have a trend that decreases with time, especially in early time. (2) PreCD4 has a positive influence on CD4 cell percentage, but it gradually becomes weak with time. (3) Age and smoking status are less influence on the CD4 percentage.  These results are quite similar to those of Huang {\it et al.} (2004).  In addition we want to note that the linear splines enable us to understand the fluctuation of coefficient functions more clearly.

\section{Concluding Remarks}
In this article, we have developed a varying-coefficient model based on basis expansion approach by maximum penalized likelihood procedure.
In order to choose values of regularization parameters, we have introduced model selection criteria from the information-theoretic and Bayesian viewpoints. 
We have applied our proposed method into some synthetic examples and Multicenter AIDS Cohort Study data. These results offers the effectiveness of our modeling strategy.  Due to the stability and the predictive performance of the constructed models, our varying-coefficient modeling strategy has the potential to be useful in a variety of practical applications.

In the future work, we will extend our model to discrete response for longitudinal data by using generalized linear models.

\section*{Appendix}
\subsection*{A.1  Derivation of the GBIC}
We consider the prior density of $\bm\theta$ as
\begin{align}
\pi(\bm\theta|\lambda_1,\ldots, \lambda_p) = 
\prod_{k=1}^{p}(2\pi)^{-\frac{M_k - r_k}{2}}(n\lambda_k)^{\frac{M_k - r_k}{2}}|\Omega_k|^{\frac{1}{2}}\exp\left\{-\frac{n}{2}\lambda_k\bm\gamma_k\Omega_j\bm\gamma_j\right\},
\end{align}
where $r_k = M_k - {\rm rank}(\Omega_k)$.  Then the marginal likelihood function of $Y$, given regularization parameters $\lambda_1,$ $\ldots,$ $\lambda_p$, is given by
\begin{align*}
p(Y | \lambda_1, \ldots, \lambda_p) &= \int f(Y|\bm\theta)\pi(\bm\theta|\lambda_1,\ldots, \lambda_p)d\bm\theta\\
& = \int \exp\left[n \times \frac{1}{n}\log\{f(Y|\bm\theta)\pi(\bm\theta|\lambda_1,\ldots, \lambda_p)\}\right]d\bm\theta\\
&\approx \frac{(2\pi)^d}{n^{d/2}|R(\hat{\bm\theta})|^{1/2}}\exp\left[n \times \frac{1}{n}\log\{f(Y|\hat{\bm\theta})\pi(\hat{\bm\theta}|\lambda_1,\ldots, \lambda_p)\}\right],
\end{align*}
where $d=\sum_{k=1}^p m_k + 1$, and the Laplace approximation is applied. Multiplying minus twice of the marginal log-likelihood function, we have
\begin{align*}
-2\log p(Y|\lambda_1, \ldots, \lambda_p) \approx & -2f(Y|\hat{\bm\theta}) + n\sum_{k=1}^{p} \lambda_k\hat{\bm\gamma}'_k\Omega_k\hat{\bm\gamma}_k - \left(\sum_{k=1}^p r_k + 1\right)\log(2\pi) \\ 
&+\left(\sum_{k=1}^p r_k + 1\right)\log n 
- \sum_{k=1}^p \log|\Omega_k| + \log|R(\hat{\bm\theta})|.
\end{align*}

\subsection*{A.2  Elements of matrices of the GIC}
Elements Matrices in GIC, defined in (\ref{RQ}), are given as follows: 
\begin{align*}
&\sum_{i=1}^{n}\left.\frac{\partial^2 l_\lambda^{(i)}(\bm\theta)}{\partial\bm\gamma_k\partial\bm\gamma_l'}\right|_{\hat{\bm\theta}} = 
\left\{\begin{array}{l}
\displaystyle{-\sum_{i=1}^n\Phi_{ik}D_{ik}\hat\sigma^{-2} S_i^{-1}D_{ik}\Phi_{ik} - n\lambda_k\Omega_k,  (k = l)}\\
\displaystyle{ -\sum_{i=1}^n\Phi_{ik}D_{ik}\hat\sigma^{-2} S_i^{-1}D_{il}\Phi_{il}, (k\neq l)}
\end{array}\right.\\
&\sum_{i=1}^{n}\left.\frac{\partial^2 l_\lambda^{(i)}(\bm\theta)}{\partial\bm\gamma_k\partial\sigma^2}\right|_{\hat{\bm\theta}} = -\frac{1}{\hat\sigma^4}\sum_{i=1}^n \Phi_{ik}D_{ik}S_i^{-1}(\bm y_i - \sum_{r=0}^p D_{ir}\Phi_{ir}\hat{\bm\gamma}_r),\\
&\sum_{i=1}^{n}\left.\frac{\partial^2 l_\lambda^{(i)}(\bm\theta)}{\partial\sigma^2\partial\bm\gamma_k}\right|_{\hat{\bm\theta}} = \left(\sum_{i=1}^{n}\left.\frac{\partial^2 l_\lambda^{(i)}(\bm\theta)}{\partial\bm\gamma_k\partial\sigma^2}\right|_{\hat{\bm\theta}}\right)',\\
&\sum_{i=1}^{n}\left.\frac{\partial^2 l_\lambda^{(i)}(\bm\theta)}{\partial\sigma^2\partial\sigma^2}\right|_{\hat{\bm\theta}} = -\frac{n}{2\hat\sigma^4},\\
\end{align*}
\begin{align*}
&\sum_{i=1}^{n}\left.\frac{\partial l_\lambda^{(i)}(\bm{\theta})}{\partial\bm{\gamma}_k}\frac{\partial l^{(i)}(\bm{\theta})}{\partial\bm{\gamma}_l'}\right|_{\hat{\bm\theta}} = \frac{1}{\hat\sigma^4}\sum_{i=1}^n \Phi_{ik}D_{ik}S_i^{-1}\Lambda_i^2S_i^{-1}D_{il}\Phi_{il} - \frac{1}{\hat\sigma^2}\lambda_k\Omega_k\hat{\bm\gamma}_k\sum_{i=1}^n\bm 1_{n_i}\hat\Lambda_iS_i^{-1}D_{il}\Phi_{il},\\
&\sum_{i=1}^{n}\left.\frac{\partial l_\lambda^{(i)}(\bm{\theta})}{\partial\bm{\bm{\gamma}_k}}\frac{\partial l^{(i)}(\bm{\theta})}{\partial\sigma^2}\right|_{\hat{\bm\theta}} = - \frac{1}{2\hat\sigma^4}\sum_{i=1}^n \Phi_{ik}'D_{ik}S_i^{-1}\hat\Lambda_i\bm 1_{n_i} + \frac{1}{2\hat\sigma^6}\Phi_{ik}'D_{ik}S_i^{-1}\hat\Lambda_i^3 S_i^{-1}\bm 1_{n_i}\\
&\sum_{i=1}^{n}\left.\frac{\partial l_\lambda^{(i)}(\bm{\theta})}{\partial\sigma^2}\frac{\partial l^{(i)}(\bm{\theta})}{\partial\bm{\bm{\gamma}_k}}\right|_{\hat{\bm\theta}} = \left(\sum_{i=1}^{n}\left.\frac{\partial l_\lambda^{(i)}(\bm{\theta})}{\partial\bm{\bm{\gamma}_k}}\frac{\partial l^{(i)}(\bm{\theta})}{\partial\sigma^2}\right|_{\hat{\bm\theta}}\right)'\\
&\sum_{i=1}^{n}\left.\frac{\partial l_\lambda^{(i)}(\bm{\theta})}{\partial\sigma^2}\frac{\partial l^{(i)}(\bm{\theta})}{\partial\sigma^2}\right|_{\hat{\bm\theta}} = 
\frac{1}{4\hat\sigma^8}\sum_{i=1}^n\bm 1_{n_i}'S_i^{-1}\hat\Lambda_i^4S_i^{-1}\bm 1_{n_i} - \frac{n}{4\hat\sigma^4},
\end{align*}
where $\bm 1_n = (1,\ldots, 1)'$ is an $n$-dimensional vector and $\Lambda_i = {\rm diag}\{y_{i1} - \sum_{k=0}^p x_{i1k}\hat{\bm\gamma}'_k\bm\phi^{(k)}(t_{i1}),$ $\ldots,$ $y_{in_i} - \sum_{k=0}^p x_{in_ik}\hat{\bm\gamma}'_k\bm\phi^{(k)}(t_{in_i})\}$.

\end{document}